\documentclass[twocolumn]{aastex63}

\usepackage{textcomp}

\newcommand{\spi}{{Spitzer}}
\newcommand{\wise}{{WISE}}

\newcommand{\iso}{{ISO}}

\providecommand{\mum}{\mbox{$\mu$m}}

\newcommand{\bcd}{BCD}
\newcommand{\tint}{\mbox{$t_{\mathrm {int}}$}}
\newcommand{\fnu}{\mbox{$F_{\nu}$}}
\newcommand{\fione}{\mbox{$F_{3.6}$}}
\newcommand{\fitwo}{\mbox{$F_{4.5}$}}

\submitjournal{AJ}\received{October 27, 2021}
\revised{\today}

\shorttitle{IRAC Observations of IRS Calibration Stars} 
\shortauthors{Kraemer et al.}

\begin{document}

\title{Tying Spitzer's IRS Calibration to IRAC: Observations of IRS Standard Stars}

\correspondingauthor{Kathleen E. Kraemer}
\email{kathleen.kraemer@bc.edu}

\author[0000-0002-2626-7155]{Kathleen E. Kraemer}
\affiliation{Institute for Scientific Research, Boston College,\\
140 Commonwealth Avenue, Chestnut Hill, MA 02467, USA}

\author{Charles W. Engelke} 
\affiliation{Institute for Scientific Research, Boston College,\\
140 Commonwealth Avenue, Chestnut Hill, MA 02467, USA}

\author{Bailey A. Renger} 
\affiliation{Institute for Scientific Research, Boston College,\\
140 Commonwealth Avenue, Chestnut Hill, MA 02467, USA}

\author[0000-0003-4520-1044]{G. C. Sloan}
\affiliation{Space Telescope Science Institute, 3700 San Martin
Drive, Baltimore, MD 21218, USA}
\affiliation{Department of Physics and Astronomy, University of
  North Carolina, Chapel Hill, NC 27599-3255, USA}

\begin{abstract}
We present 3.6 and 4.5~\mum\ photometry for a set of 61 standard stars 
observed by \spi's Infrared Spectrograph (IRS).  The photometry was obtained
with the Infrared Array Camera (IRAC) on \spi\ in order to help tie the 
calibration of IRAC and the IRS, which had been anchored to the calibration 
of the Multiband Infrared Photometer for \spi\ (MIPS).  The wavelength 
range of the IRS data only slightly overlaps with the IRAC 4.5~\mum\ band and 
not at all with the 3.6~\mum\ band.  Therefore, we generated synthetic spectra 
from spectral templates of stars with the same spectral types and luminosity 
classes as our sample stars, normalized to the IRS data at 6--7 \mum, and 
compared those to the observed photometry.  
The new IRAC observations of  IRS standard stars demonstrate that the two instruments are calibrated to within 1\% of each other.
\end{abstract}

\section{Introduction\label{sec.intro}}

The calibration of the Infrared Spectrograph \cite[IRS;][]{irs04} on the \spi\ 
{Space Telescope} is based on numerous observations of spectrophotometric 
standard stars, combined with model atmospheres \citep{irscal15}. The two 
primary standards were the A dwarfs $\alpha$ Lac and $\delta$ UMi, and 
the calibration was then transferred to the rest of the A-dwarf and K-giant 
sample. The absolute flux calibration was pinned at the red end of the 
spectrum to the 24 \mum\ calibration of the Multiband Imaging Photometer for 
\spi\ \cite[MIPS;][]{mips04, riekeea08} using 22 \mum\ photometry from the 
IRS Red Peak-Up array. \cite{irstr12003} estimated that the uncertainty in the 
spectroscopic calibration, that is, the point-to-point flux densities 
of the spectra, is better than 1\%.

\begin{figure} 
\begin{minipage}{\linewidth}
\includegraphics[width=0.99\textwidth]{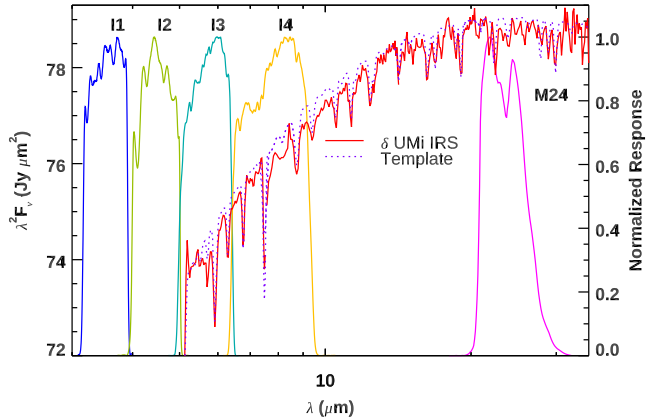}
\caption{IRS spectrum of $\delta$ UMi (red line) compared to a scaled Kurucz
model (purple dotted line). The spectra are shown in Rayleigh-Jeans (RJ) flux
units, $\lambda^2F_{\nu}$, in which the RJ tail of a blackbody would be
a horizontal line. Also shown are the IRAC bands and the MIPS 24
\mum\ band, each normalized to a peak of 1.}
\label{fig.irsirac}
\end{minipage}
\end{figure}

However, this estimate does not include the uncertainty in the absolute flux calibration of the IRS. In particular,
the short-wavelength end of the spectra was never formally compared 
to photometric measurements during the cryogenic mission. Ideally, the 
calibration stars would have been observed with the Infrared Array Camera 
\cite[IRAC;][]{irac04} I3 and I4 bands at 5.8 and 8.0 \mum\ concurrently with 
IRS observations, as those bands overlapped with the IRS spectral coverage 
(Figure \ref{fig.irsirac}). This would have provided the most straightforward 
cross-calibration between the two instruments, and could have helped verify 
the 1.5\% difference between MIPS and IRAC reported by \cite{riekeea08}. 

Since such standards were not observed  during the cryogenic 
phase, we obtained Warm \spi\ observations in the 3.6 
and 4.5~\mum\ bands (I1 and I2).  Section 2 describes the observations 
and data 
processing.  We compare the measured photometry to synthetic photometry from 
stellar templates for each star in Section 3, and 
Section 4 summarizes our findings.

\section{Observations and Data Reduction\label{sec.obs}} 

\subsection{Source Selection and IRS Data} 

The IRS instrument team observed several dozen calibration sources over the 
course of the mission, primarily K giants and A dwarfs, along with bright 
stars and very red sources such as galaxies.  We selected the set of
K giants and A dwarfs described by \citet{irscal15}. We also included several 
fainter K giants and solar analogs that were added to the calibration list 
later in the mission.  Our sample contained 61 stars:  20 A dwarfs, 3 G 
dwarfs (solar analogs), 
and 38 K giants. To simplify the discussion, we include the one B9 dwarf and two late G giants in the A dwarf and K giant categories, respectively.
Tables \ref{tab.starsa} and \ref{tab.starsk} list the observed sample, with
the solar analogs included with the A dwarfs in Table \ref{tab.starsa}.

For the IRS spectra of these sources, we use the spectral data reductions from \citet{irscal15}, which readers should refer to for details on the IRS data processing. The spectral types listed in Tables \ref{tab.starsa} and 
\ref{tab.starsk} are also taken from \citet{irscal15}, with two exceptions.
We assign a spectral type of A1 V for HD 165459, following 
\citet{reachea05}. The spectral type for $\lambda$ Tel is discussed 
in Section \ref{sec.templ}.

\subsection{IRAC Observations} 

Each star was observed in subarray mode with the ``Small 4-Position Gaussian'' 
dither pattern to mitigate bad pixels and gain variations.  
The integration times (\tint) were based on the flux densities (\fnu) 
estimated from the partially saturated measurements 
from the Wide-field Infrared Survey Experiment \citep[\wise;][]{wise10}
in the AllWISE source catalog \citep{cutriea13}.  Although the 
saturation compromised the data for calibration use, they could still be 
used for estimating IRAC flux densities and setting integration times.  
These were \tint= 0.02 sec ($F_\nu\!>$3 Jy), 0.1 sec (0.7--3 Jy), 
or 0.4 sec ($<$0.7 Jy), i.e., the longest time without saturating.  Both 
bands used the same integration time to simplify the observational set up. 
With the chosen integration 
times and estimated flux densities, the expected signal-to-noise ratios were
$\sim$100--200 in each band for all stars. 
Table \ref{tab.starsa} includes observing details for the A dwarfs and Table 
\ref{tab.starsk} gives those for the K giants. 

Most of the observations were made under Program 12059 in 2015--2016, with 
11 additional stars observed under Program 14221 in 2019.
Thirteen stars were observed twice because the temperature of the 4.5~\mum\ 
array was unstable due to an IRAC/Spacecraft Communications anomaly from 2015 November 27 to December 14, the time of the first observations 
(S. Carey, 2016 
Jan 14, private communication).  As explained in Section~\ref{sec.tempcomp},
we include both observations for these stars.  HD 165459, a calibration 
target for IRAC, was observed three 
times, once at each integration time.  Our final sample contains 76 
observations of the 61 stars. 

\begin{deluxetable*}{lllllll}
\tablecaption{Observations: A Dwarfs \& Solar Analogs\label{tab.starsa}}
\tablewidth{0pt}
\tablehead{
\colhead{Star}  &\colhead{RA} &\colhead{Dec} & \colhead{Spectral} &
\colhead{Observ.} &\colhead{AOR} &  \colhead{\tint} \\
\colhead{Name}  &\colhead{(J2000)} &\colhead{(J2000)}&  \colhead{Type} & 
\colhead{Date} & \colhead{Key} & \colhead{(sec)}
}
\startdata
$\lambda$ Tel &18 58 27.77  
&$-$52 56 19.1 & B9 IV/V\tablenotemark{a} & 2015 Dec 16 & 58140160 & 0.02 \\
\nodata\tablenotemark{b}       &    \nodata                             
&   \nodata               &  \nodata       & 2016 Jun 27 & 58898176 & 0.02 \\   
$\eta^1$ Dor  &06 06 09.38
&$-$66 02 22.6 & A0 V & 2019 May 04  & 68758016 & 0.02  \\
HD 46190      &06 27 48.62  
&$-$62 08 59.7 & A0 V & 2016 Jan 20 & 58141184 & 0.10   \\
$\xi^1$ Cen   &13 03 33.31  
&$-$49 31 38.2 & A0 V & 2016 May 12 & 58139904 & 0.02   \\
\enddata
\tablenotetext{a}{See text regarding the luminosity class of $\lambda$ Tel.}
\tablenotetext{b}{Ellipses indicate $\lambda$ Tel was one of the stars 
observed twice.}
\tablecomments{This table is available in its entirety in machine-readable 
format. 
}
\end{deluxetable*}

\begin{deluxetable*}{lllllll}
\tablecaption{Observations: K Giants\label{tab.starsk}}
\tablewidth{0pt}
\tablehead{
\colhead{Star}  &\colhead{RA} &\colhead{Dec} & \colhead{Spectral} &
\colhead{Observ.} &\colhead{AOR} &  \colhead{\tint} \\
\colhead{Name}  &\colhead{(J2000)} &\colhead{(J2000)}&  \colhead{Type} &
\colhead{Date} & \colhead{Key} & \colhead{(sec)}
}
\startdata
BD+00 2862   &11 56 54.19 &$-$00 30 13.5 
& G8 III & 2016 Mar 20 & 58130432 & 0.40 \\
HR 6606      &17 37 08.88 &  +72 27 20.9
& G9 III & 2019 Apr 01 & 68757504 & 0.02\\
HD 41371     &06 00 07.71 &$-$64 18 36.0
& K0 III & 2019 Apr 25 & 68756992 & 0.02\\
HD 51211     &06 50 25.27 &$-$69 59 10.5 
& K0 III & 2015 Dec 18 & 58136576 & 0.02 \\
   \nodata  &       \nodata      &      \nodata                         
&  \nodata      & 2016 May 26 & 58898944 & 0.02 \\
\enddata 
\tablecomments{This table is available in its entirety in machine-readable 
format. }
\end{deluxetable*}

\subsection{Data Reduction\label{sec.datared}} 

The processing pipeline started with the basic calibrated data (\bcd) from 
the \spi\ archive, pipeline version 19.2.0. In the IRAC subarray mode, an 
observation consists of 64 frames, which are provided in the \bcd\ fits file. 
Each \bcd\ corresponds to one of the four dither positions, and each star was 
observed in both the I1 and I2 bands.

Aperture photometry was performed on each frame with the {\em sextractor} 
software 
 with a 6-pixel diameter for the aperture and a (local) background 
annulus of 8 pixels. To correct for photometric variations due to 
the location of the source on the
array and within a pixel, we used the IRAC\_APHOT\_CORR.pro\footnote{https://irsa.ipac.caltech.edu/data/SPITZER/docs/
dataanalysistools/tools/contributed/}
 created by J. Ingalls and provided by the \spi\ Science Center (SSC),
using positions from the SSC's BOX\_CENTROIDER procedure. These
corrections to the intrapixel response can be up to a few percent, so they are
important to take into account for this project.
{\em Sextractor} positions are determined by a different method, though, and 
cannot be used directly with IRAC\_APHOT\_CORR.

The aperture corrections were based on the values in the IRAC Handbook 
(v4.0), Table 4.2.  The Handbook gives the conversion factors 
{\em zmag} and $C$ 
to convert from surface brightness to flux density for 0\farcs6 pixels 
(appropriate for mosaics from, e.g., MOPEX), as opposed to the 1\farcs2 native 
pixels used here. For the larger pixels, the conversion factor becomes
$C$=33.84638$\times10^{-6}$ Jy/pixel/(MJy/sr) (cf. 8.461595$\times10^{-6}$ 
Jy/pixel/(MJy/sr) for the 
0\farcs6 pixels) and {\em zmag}=17.297 mag for I1 and 16.813 mag for I2.

The photometry from the 64 frames was averaged into a single value for each
dither position (i.e., each \bcd). The standard deviation among the frames 
of a given position was {$\sigma_f\!\sim$0.3--2.3\% and 0.4--3.4\% with
means of 0.7\% and 1.0\% for I1 and I2, respectively. The variation among 
the four \bcd s for a given star  $\sigma_b\!\sim$0.2--1.5\% and 0.2-2.0\%
with means of 0.8\% for both I1 and I2.

\subsection{Photometric Calibration with HD 165459\label{sec.cal}}  

We used archival IRAC data of HD 165459, which 
was one of the primary IRAC calibration sources \citep[e.g.,][]{reachea05}, 
to calibrate our observations and verify our processing pipeline.
\citet{reachea05} give the spectral type of HD 165459 as A1 V.  They also
give its brightness as [3.6] = 6.593 mag  and [4.5] = 6.575 mag, which 
correspond to \fione = 0.648 Jy and \fitwo=0.421 Jy. 
We use these values as
the fiducial standard values in the following discussion.
  
We processed 22 IRAC calibration observations taken with the subarray
between 30 Oct 2015 and 17 Sep 2016, which span the observation
dates for  PID 12059. These observations used the large-scale, 4-position, 
 Gaussian dither pattern with \tint=0.4 sec (the two calibration
observations with the subarray at 0.02 or 0.1 sec in this time frame were not 
dithered). 

The result of processing these data finds that the flux densities
from our pipeline were too faint compared to \citet{reachea05} by 
3.64 $\pm$ 0.21\% for I1 and 3.57 $\pm$ 
0.27\% for I2. The uncertainties, $\sigma_{cal}$, are the standard deviations 
among the 88 measurements (4 dither positions for each of the 22 observations).
Additional details, including the photometry, are given in the 
Appendix.

Table \ref{tab.flux} reports the resulting flux densities for our targets. 
These are the weighted means of the four \bcd s (weighted by $\sigma_f$), 
adjusted in magnitude space by subtracting the magnitude offsets 
from the HD 165459 
calibration data to correct the 
bias in our pipeline. This aligns our results with the photometric
calibration of IRAC by \citet{reachea05}. The quoted uncertainties 
are from $\sigma_b$, the standard
deviation of those four measurements, and $\sigma_{cal}$, the calibration bias
uncertainty, added in quadrature.

\startlongtable
\begin{deluxetable}{llRR}
\tablecaption{Observed Photometry\label{tab.flux}}
\tablewidth{0pt}
\tablehead{
\colhead{Star}  &\colhead{AOR} & \colhead{\fione} &
\colhead{\fitwo}   \\
\colhead{Name}  &\colhead{Key} &\colhead{(Jy)}&  \colhead{(Jy)}
}
\startdata
  $\lambda$ Tel & 58140160 &   3.078 $\pm$   0.034 &   1.960 $\pm$   0.003 \\
        \nodata & 58898176 &   3.070 $\pm$   0.018 &   1.959 $\pm$   0.011 \\
   $\eta^1$ Dor & 68758016 &   1.405 $\pm$   0.009 &   0.891 $\pm$   0.007 \\
       HD 46190 & 58141184 &   0.799 $\pm$   0.005 &   0.512 $\pm$   0.004 \\
    $\xi^1$ Cen & 58139904 &   3.586 $\pm$   0.034 &   2.296 $\pm$   0.018
\enddata
\tablecomments{This table is available in its entirety in machine-readable
format. 
}
\end{deluxetable}

\subsection{Array Temperature Drift Effect\label{sec.tempcomp}} 

Thirteen stars were observed twice due to a concern by the SSC regarding a
small drift in the temperature of the I2 focal plane array (FPA) from
an IRAC/Spacecraft anomaly. The mean temperature of the 
array was $T_{\mathrm{4.5,~v1}} = 27.873 \pm 0.018$ K for the first set (those 
taken before HMJD 57375) and $T_{\mathrm{4.5,~v2}} = 28.687 \pm 0.003$ K for 
the second 
set; quoted uncertainties are 1-$\sigma$ standard deviations. For comparison, 
the temperatures at 3.6~\mum\ were $T_{3.6} = 28.655 \pm 0.001$ K and 
$28.708 \pm 0.002$ K, respectively. Figure \ref{fig.fpar} 
shows the ratios of the first to second measurements $R$ (v1/v2) = 
$F_\nu$ (v1)/$F_\nu$ (v2) as a function of the flux density from the first 
observation, \fione (v1).  

The mean ratios are 1.002 $\pm$ 0.001 and 
1.003 $\pm$ 0.001 for bands I1 and I2, respectively. That is, the second
measurements are slightly lower than the first measurements, although the
statistical significance is marginal.
For individual stars, the difference between the two observations 
is always within the uncertainty of at least one measurement and 
most are 
them for both measurements (22 of the 26 data points).
We conclude that the change in FPA temperature for I2 does not affect the 
measured photometry and therefore include both measurements in our analysis.

\begin{figure} 
\includegraphics[width=3.35in]{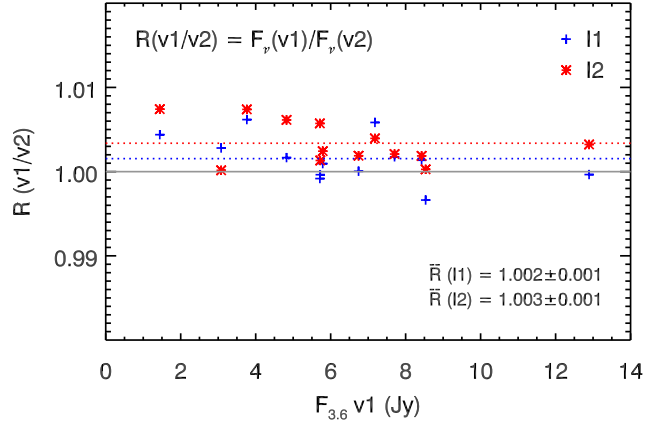}
\caption{The ratios of the first and second measurements (v1 and v2,
respectively) for I1 (blue pluses) and I2 (red asterisks) for the 13 stars
with two observations.  The flux density ratios are plotted vs.\ \fione\ v1.
The blue dotted line shows the mean ratio for I1 and the 
red dotted line shows the mean ratio for I2.
\label{fig.fpar}
}
\end{figure}

\subsection{Spectral Templates\label{sec.templ}}  

\begin{figure*}[h!]  
\includegraphics[width=0.9\textwidth]{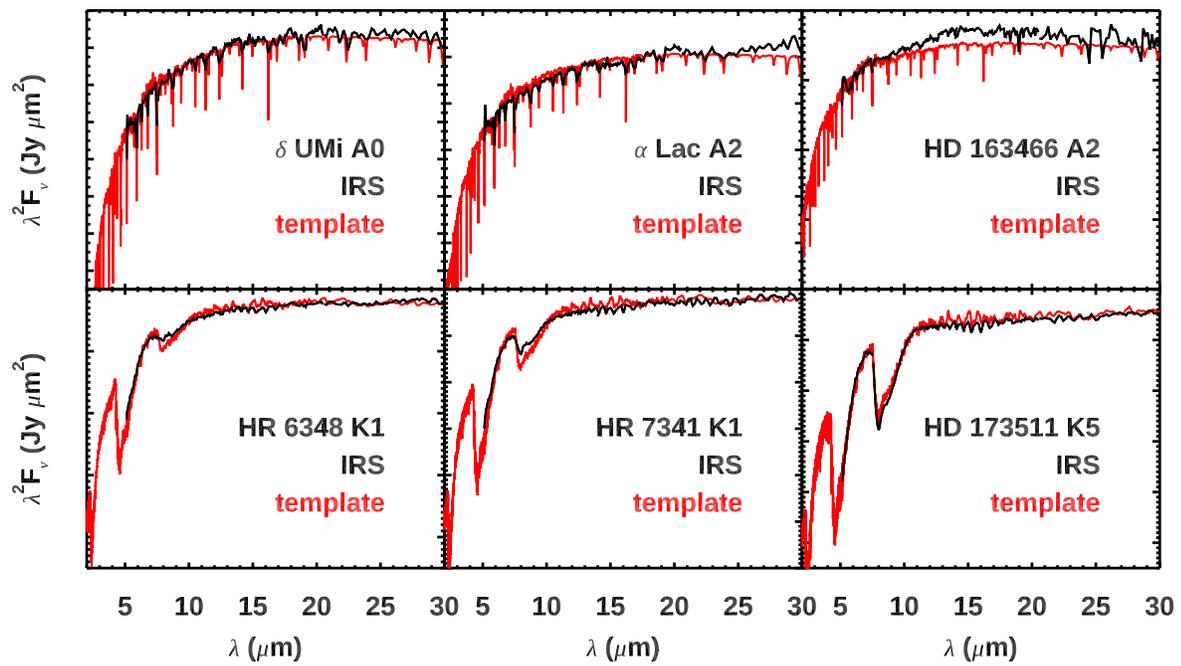}
\caption{Comparison of IRS data and spectral templates for six of the
IRS standards observed the most frequently, in RJ flux units. Templates (red) 
were normalized to the IRS data (black) between 6 and 7 \mum. 
}
\label{fig.spectra}
\end{figure*}
     
The IRS wavelength range did not significantly overlap with the I1 and I2 
bands, so we cannot
compare the IRAC photometry to the IRS spectra directly.
To extend the spectra to shorter wavelengths and generate synthetic 
photometry, we instead use the stellar templating procedures pioneered by
M. Cohen and colleagues \cite[e.g.,][]{cohenea92ii, cohenea95iv, cohenea99x} 
and updated by \cite{engelkeea06}.  The templates for the K giants are 
derived from the stars of \cite{engelkeea06}, which were based on the 
spectral library of \cite{pickles98} and the spectral atlas of \cite{swsatlas}
from the Short-Wavelength Spectrometer \citep[SWS;][]{sws96} on the 
\textit{Infrared Space Observatory} \citep[\iso;][]{iso96}.  The templates
for the A dwarfs are based on Kurucz atmospheric
models.\footnote{http://kurucz.harvard.edu/}  The templates are 
interpolated in temperature space and angular diameter to fit the individual 
stars in our sample. The spectral type is used to set the initial temperature
 and diameter, and a least-squares fit to visible and infrared photometry 
from the literature is performed.

Each template is normalized to the corresponding IRS data between 6 and 
7 \mum, i.e., between the molecular absorption bands dominated by CO at 5~\mum\
and SiO starting at 7.5~\mum\ in the K giants.  This range was also used for 
the A dwarfs for consistency.  Figure \ref{fig.spectra} shows the
 IRS data for a set of the most frequently observed A dwarfs and K giants, 
which thus have high signal-to-noise ratios, along with the 
corresponding spectral templates used for the synthetic photometry.

Seven stars were not templated, although their IRAC photometry is included
in the tables.  HD 162317 has a very noisy IRS spectrum which precludes
a reliable normalization for the template. $\lambda$ Tel has an ambiguous
spectral type suggesting that it is not a normal main-sequence 
star.\footnote{\citet{irscal15} report a spectral type of B9 III, based on 
\citet{devauc57} and \citet{bm58}, but it is classified as A0 III* by
\citet{gg87} and B9.5 IV/V by \citet{mich75vol1}.  \citet{soubiranea16} 
give log $g$=3.42, more consistent with a dwarf than a giant.}
HD 196850, HD 176841, HD 73350 are solar analogs, but the IRS spectra 
for two of them were too noisy to use. Also, we do not have a full set of 
validated templates for G dwarfs and did not have a good match for the third.  
HR 5949 and HD 46190 have excess 
emission from debris disks even at the shorter wavelengths 
\citep[cf.,][their Fig.\ 24]{irscal15}.

Overall, we have 16 templated A dwarfs with 18 IRAC observations and 
37 templated K giants with 47 observations. 

\section{Results}  

We generated the synthetic photometry from the templates using the 
subarray-specific relative spectral response functions (RSRFs) for the warm 
mission obtained from the SSC 
website\footnote{https://irsa.ipac.caltech.edu/data/SPITZER/docs/irac/ calibrationfiles/spectralresponse/}. 
Figure \ref{fig.rsrf} shows the RSRFs, two sample templates, and their IRS 
spectra, along with the normalization region.  Table \ref{tab.ratios} gives 
the synthetic and observed flux densities  and their ratio in the two bands
for each star with a template.

\begin{figure}  
\begin{minipage}{\linewidth}
\includegraphics[width=0.99\textwidth]{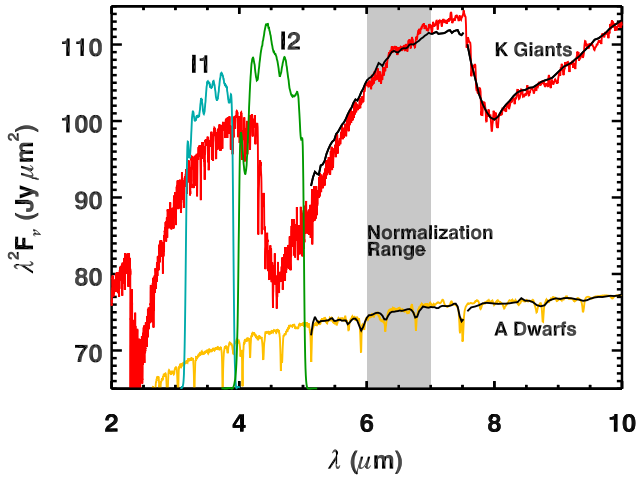}
\caption{IRAC subarray RSRFs  with representative spectra for a K giant 
and an A dwarf. Band I1 (3.6 \mum) is in cyan and I2 (4.5 \mum) in
green. The black curves show measured IRS data for two stars in our sample,
and the red and gold curves are the SWS-based templates. The templates were
normalized to the IRS data at 6.0--7.0~\mum\ (gray box).}
\label{fig.rsrf}
\end{minipage}
\end{figure}

Figure \ref{fig.comp} shows the results for each band, with the A  dwarfs 
and K giants shown separately.  The top section of each panel compares
the IRAC flux densities to the synthetic photometry, and the lower section 
shows the ratios.  

\begin{figure*}  
\includegraphics[width=7.in]{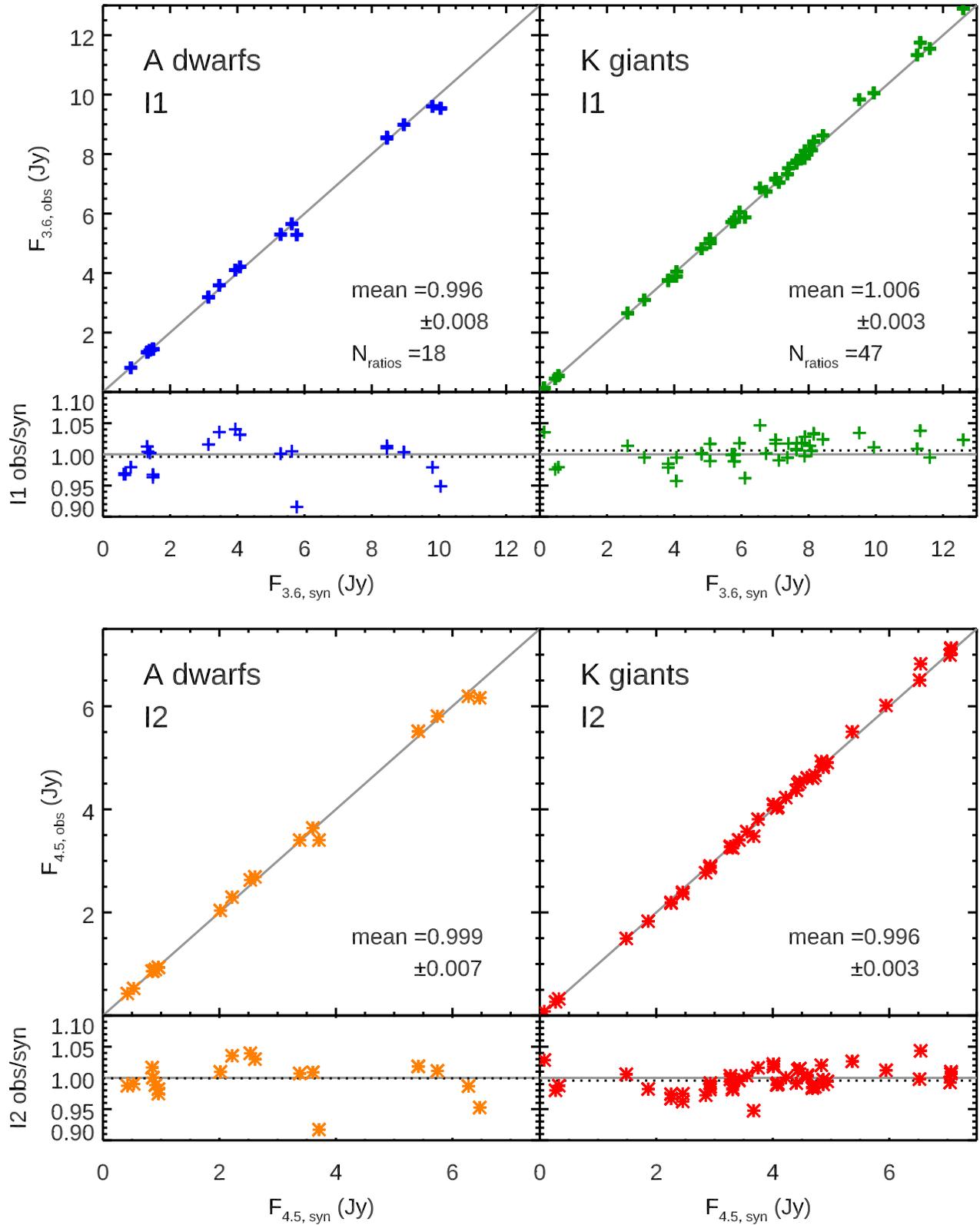}
\caption{Observed and synthetic photometry comparison. The upper portion of 
each panels shows the synthetic photometry as a function of the observed 
photometry. The lower portions show the ratio of observed to synthetic 
photometry. The solid gray lines indicate equality between the two sets of 
photometry.
(left) The A star panels include 18 observations of 16 stars. (right) The K 
stars include 47 observations of 37 stars. I1 and I2 are in the top and bottom
panels, respectively. The dotted lines in ratio panels show the mean 
for that subset.
}
\label{fig.comp}
\end{figure*}

\begin{deluxetable*}{lRRRRRR}
\tablecaption{Observed to Synthetic Ratios\label{tab.ratios}}
\tablewidth{0pt}
\tablehead{
\colhead{} & \multicolumn{3}{c}{I1}& \multicolumn{3}{c}{I2}\\
\colhead{Star} & \colhead{\fione$_{obs}$} & \colhead{\fione$_{syn}$} &
\colhead{Obs/Syn} & \colhead{\fitwo$_{obs}$} & \colhead{\fitwo$_{syn}$} &
\colhead{Obs/Syn}\\
\colhead{Name} & \colhead{(Jy)} & \colhead{(Jy)} & \colhead{ } 
            & \colhead{(Jy)} & \colhead{(Jy)} & \colhead{ } 
}
\startdata
 $\alpha$ Lac    &   8.986 &   8.955 &   1.004 &   5.807 &   5.743 &   1.011 \\
 $\delta$ UMi    &   5.650 &   5.622 &   1.005 &   3.637 &   3.605 &   1.009 \\
$\epsilon$ Aqr   &   9.534 &  10.049 &   0.949 &   6.163 &   6.470 &   0.953 \\
    $\mu$ PsA    &   5.294 &   5.289 &   1.001 &   3.403 &   3.380 &   1.007 \\
    $\nu$ Tau v1 &   8.533 &   8.451 &   1.010 &   5.515 &   5.415 &   1.018 \\
    $\nu$ Tau v2 &   8.562 &   8.451 &   1.013 &   5.513 &   5.415 &   1.018 
\enddata
\tablecomments{This table is available in its entirety in machine-readable
format. Labels ``v1'' and ``v2'' in the names refer to photometry from the
first and second measurements for the 13 stars which were observed twice.
}
\end{deluxetable*}

\subsection{A Dwarfs}  

The 18 observations of the 16 A dwarfs show very good agreement between the
observed and synthetic photometry.  They have a mean ratio of observed 
IRAC to synthetic photometry of 0.996 $\pm$ 0.008 at 3.6~\mum\ and 0.999
$\pm$ 0.007 at 4.5~\mum.  The uncertainties here are the uncertainties
in the mean, i.e., the standard deviation divided by $\sqrt{N}$ where N=18.

We can also compare our results to the recent analysis of \cite{krickea21}.
They analyzed the IRAC photometry of stars which may be used for 
calibration of the \textit{James Webb Space Telescope} (JWST), from both 
the cryogenic and post-cryogenic portions of the \spi\ mission.
Table \ref{tab.krick} compares the photometry for the five sources we have in 
common.  
The mean ratio of our photometry to that of Krick et al.\
is 1.002 at 3.6~\mum\ and 0.995 at 4.5~\mum, respectively. For both wavelengths,
 the uncertainties in the mean are 0.003.} The individual ratios 
are also statistically indistinguishable from unity.
We conclude that our processing agrees with the results of 
\cite{krickea21}, which were based on a much larger set of observations 
taken with a wide range of observing settings.

\subsection{K Giants}  

For the 47 observations of the 37 K giants, the mean ratios of the
observed IRAC photometry to the synthetic photometry are 1.006 $\pm$
0.003 at 3.6~\mum\ and 0.996 $\pm$ 0.003 at 4.5~\mum.  The 1--2 $\sigma$
deviation from unity may reflect variations in the strengths of the molecular 
absorption bands present in individual K giants, as has been seen previously 
\citep[e.g.,][]{herasea02, engelkeea06}.
As with the A 
dwarfs, the IRAC and IRS data agree well at both wavelengths.

\begin{deluxetable*}{lllll}
\tablecaption{Krick et al. Photometry Comparison\label{tab.krick}}
\tablewidth{0pt}
\tablehead{
\colhead{Star} & \multicolumn{2}{c}{\fione\ (Jy)} & 
\multicolumn{2}{c}{\fitwo\ (Jy)}\\
\colhead{Name} & \colhead{This work} & \colhead{Krick+21} & 
\colhead{This work} & \colhead{Krick+21}
}
\startdata
$\eta^1$ Dor & 1.405$\pm$0.009 & 1.395$\pm$0.002 & 0.862$\pm$0.007 & 0.887$\pm$0.002\\
HR 5467      & 1.334$\pm$0.012 & 1.335$\pm$0.001 & 0.860$\pm$0.004 & 0.866$\pm$0.001\\
$\delta$ UMi & 5.765$\pm$0.023 & 5.679$\pm$0.001 & 3.637$\pm$0.023 & 3.658$\pm$0.001\\
HD 163466    & 0.813$\pm$0.009 & 0.816$\pm$0.001 & 0.525$\pm$0.005 & 0.529$\pm$0.001\\
HD 165459\tablenotemark{a} & 0.660$\pm$0.025 & 0.652$\pm$0.001 & 0.421$\pm$0.001 & 0.425$\pm$0.001
\enddata
\tablenotetext{a}{Flux densities of HD 165459 listed for this work are the means of the three observations and the uncertainty is the standard deviation among the three.}
\end{deluxetable*}

\section{Summary\label{sec.summ}}

We have presented new IRAC observations of 61 stars which were part of the IRS
calibration program. The measured photometry at 3.6 and 4.5~\mum\
was compared to synthetic photometry from stellar templates based on the 
IRS spectra normalized between 6 and 7 \mum. For the 18 observations of the 
16 A stars, the mean ratios are 0.996 $\pm$ 0.008 at 3.6~\mum\ and 0.999 
$\pm$ 0.007 at 3.6~\mum\ and  4.5~\mum, respectively. 
For the K stars, which had 47 observations of 37 stars, the mean ratios are 
1.006 $\pm$ 0.003  and 0.996 $\pm$ 0.003 at 3.6~\mum\ and  4.5~\mum, 
respectively. Overall, we conclude that the IRAC and IRS calibrations agree 
to within 1\%.

\acknowledgments
This work is based on observations made with the {\em Spitzer Space 
Telescope}, which is operated by the Jet Propulsion Laboratory, 
California Institute of Technology under NASA contract 1407.
Financial support for this work was provided by NASA through NASA ADAP 
grant NNX17AF23G.  We thank I.\ Sahinidis for performing the initial 
data quality assessment of the IRAC observations.  We made use of the
NASA Astrophysics Data System, IRSA's Gator service, and CDS's Simbad \& 
Vizier services. 
We especially thank the referee whose vital insight 
into IRAC pixel phases and source positions from different extraction
methods corrected our processing pipeline and significantly improved our 
results.

\facilities{\spi~(IRS), \spi~(IRAC)}

\appendix
%

\section{HD 165459\label{sec.app165459}}

Table \ref{tab.165459cal} gives the photometry produced by our pipeline 
for HD 165459 for
the individual calibration observations taken by the instrument team.
These used sub-array mode with the 4-position large-scale Gaussian 
dither pattern with \tint=0.4 sec. The values \fione$_{meas}$ and 
\fitwo$_{meas}$ are the means of the 4 positions 
(\bcd s), weighted by $\sigma_f$, the standard deviation among the frames 
of a given \bcd. The uncertainty reported is the weighted uncertainty.  
Also given are the differences 
between the measured value and the fiducial value from \citet{reachea05}, 
$\Delta$ from 6.593 mag 
and 6.575 mag for I1 and I2, respectively. Our pipeline gives consistently
fainter values than the fiducial in both bands. We use the weighted mean 
of the magnitude offsets to correct the photometry in magnitude space
 for our target stars in each band.

\begin{deluxetable*}{CCCRCCCCC}
\tablecaption{HD 165459 Calibration Photometry \label{tab.165459cal}}
\tablehead{
 &\colhead{[I1]} &
\colhead{\fione$_{meas}$} & \multicolumn{2}{c}{$\Delta${~from~ 6.593~mag}}& \colhead{[I2]} &
\colhead{\fitwo$_{meas}$} & \multicolumn{2}{c}{$\Delta${~from~ 6.575~mag}} \\
\colhead{AOR}  & \colhead{(mag)}  &    \colhead{(Jy)}  & \colhead{(mag)} & \colhead{(\%)}
 & \colhead{(mag)}  &    \colhead{(Jy)}  & \colhead{(mag)} & \colhead{(\%)}
}
\startdata
  57825280 &  6.625 \pm 0.002 &  0.629 \pm  0.001 &   0.032  & $-$2.9 &   6.612 \pm  0.002 &  0.407 \pm  0.001 &  0.037 & $-$3.3 \\
  57844224 &  6.628 \pm 0.002 &  0.627 \pm  0.001 &   0.035  & $-$3.3 &   6.609 \pm  0.002 &  0.408 \pm  0.001 &  0.034 & $-$3.0 \\
  58480128 &  6.626 \pm 0.002 &  0.628 \pm  0.001 &   0.033  & $-$3.1 &   6.611 \pm  0.002 &  0.408 \pm  0.001 &  0.036 & $-$3.2 \\
  58508032 &  6.628 \pm 0.002 &  0.627 \pm  0.001 &   0.035  & $-$3.3 &   6.610 \pm  0.002 &  0.408 \pm  0.001 &  0.035 & $-$3.2 \\
  58536448 &  6.629 \pm 0.002 &  0.627 \pm  0.001 &   0.036  & $-$3.2 &   6.606 \pm  0.002 &  0.409 \pm  0.001 &  0.031 & $-$2.8 \\
  58593024 &  6.629 \pm 0.002 &  0.627 \pm  0.001 &   0.036  & $-$3.2 &   6.615 \pm  0.002 &  0.406 \pm  0.001 &  0.040 & $-$3.6 \\
  58730496 &  6.629 \pm 0.002 &  0.626 \pm  0.001 &   0.036  & $-$3.3 &   6.609 \pm  0.002 &  0.408 \pm  0.001 &  0.034 & $-$3.1 \\
  58748416 &  6.631 \pm 0.002 &  0.626 \pm  0.001 &   0.038  & $-$3.5 &   6.609 \pm  0.002 &  0.408 \pm  0.001 &  0.034 & $-$3.1 \\
  58821888 &  6.632 \pm 0.002 &  0.625 \pm  0.001 &   0.039  & $-$3.6 &   6.612 \pm  0.002 &  0.407 \pm  0.001 &  0.037 & $-$3.3 \\
  58857984 &  6.634 \pm 0.002 &  0.624 \pm  0.001 &   0.041  & $-$3.7 &   6.607 \pm  0.002 &  0.409 \pm  0.001 &  0.032 & $-$2.9 \\
  58879232 &  6.628 \pm 0.002 &  0.627 \pm  0.001 &   0.035  & $-$3.2 &   6.614 \pm  0.002 &  0.406 \pm  0.001 &  0.039 & $-$3.5 \\
  58909952 &  6.631 \pm 0.002 &  0.626 \pm  0.001 &   0.038  & $-$3.5 &   6.607 \pm  0.002 &  0.409 \pm  0.001 &  0.032 & $-$2.9 \\
  58939392 &  6.627 \pm 0.002 &  0.628 \pm  0.001 &   0.034  & $-$3.1 &   6.612 \pm  0.002 &  0.407 \pm  0.001 &  0.037 & $-$3.3 \\
  58968576 &  6.632 \pm 0.002 &  0.625 \pm  0.001 &   0.039  & $-$3.6 &   6.612 \pm  0.002 &  0.407 \pm  0.001 &  0.037 & $-$3.4 \\
  59271168 &  6.631 \pm 0.002 &  0.626 \pm  0.001 &   0.038  & $-$3.4 &   6.612 \pm  0.002 &  0.407 \pm  0.001 &  0.037 & $-$3.3 \\
  59289856 &  6.627 \pm 0.002 &  0.628 \pm  0.001 &   0.034  & $-$3.1 &   6.614 \pm  0.002 &  0.407 \pm  0.001 &  0.039 & $-$3.4 \\
  59417856 &  6.628 \pm 0.002 &  0.627 \pm  0.001 &   0.035  & $-$3.3 &   6.614 \pm  0.002 &  0.407 \pm  0.001 &  0.039 & $-$3.4 \\
  59619072 &  6.630 \pm 0.002 &  0.626 \pm  0.001 &   0.037  & $-$3.4 &   6.613 \pm  0.002 &  0.407 \pm  0.001 &  0.038 & $-$3.4 \\
  60101888 &  6.629 \pm 0.002 &  0.626 \pm  0.001 &   0.036  & $-$3.3 &   6.605 \pm  0.002 &  0.410 \pm  0.001 &  0.030 & $-$2.7 \\
  60276224 &  6.631 \pm 0.002 &  0.625 \pm  0.001 &   0.038  & $-$3.5 &   6.609 \pm  0.002 &  0.408 \pm  0.001 &  0.034 & $-$3.0 \\
  60612608 &  6.631 \pm 0.002 &  0.625 \pm  0.001 &   0.038  & $-$3.5 &   6.609 \pm  0.002 &  0.408 \pm  0.001 &  0.034 & $-$3.0 \\
  60630528 &  6.631 \pm 0.002 &  0.625 \pm  0.001 &   0.038  & $-$3.5 &   6.613 \pm  0.002 &  0.407 \pm  0.001 &  0.038 & $-$3.4 \\
\enddata
\end{deluxetable*}

\newpage

\bibliographystyle{aasjournal}
\bibliography{irscalrefs1}

\end{document}